\documentclass[proceedings, preprint]{rmaa}

\usepackage{paralist}

\usepackage{psfrag,color}

\SetYear{2018}
\SetConfTitle{8\textsuperscript{th} ADeLA meeting}

\title{The GBOT Asteroid Survey\\ 
 \sl{\small{(First years: Jan. 2015 - May 2018)}}} 

\author{
  S. Bouquillon,\altaffilmark{1} 
  D. Souami,\altaffilmark{2}}

\altaffiltext{1}{SYRTE, Observatoire de Paris, PSL Research University, 
CNRS, Sorbonne Universit\'es, UPMC Univ. Paris 06, 
61 avenue de l'Observatoire, 75014, Paris, France.}

\altaffiltext{2}{LESIA, Observatoire de Paris, CNRS UMR 8109, 
Sorbonne Universit\'e, Universit\'e Paris-Diderot, 
5 place Jules Janssen, F-92195 Meudon C\'edex, France}

\shortauthor{Bouquillon, \& Souami}
\shorttitle{RevMexAA(SC) The GBOT Asteroid Survey}

\listofauthors{S. Bouquillon, \& D. Souami}
\indexauthor{Bouquillon, S.}
\indexauthor{Souami, D.}

\abstract{The GBOT group is in charge of the Ground Based Optical Tracking of the Gaia satellite. In concrete terms, since the launch of Gaia, our task is to take every night, using ground based medium-class telescopes, short sequences of $10$ or $20$ images of the Gaia satellite close to its meridian transit. 
For this purpose, we mainly use the VLT Survey Telescope and the Liverpool Telescope.
In these images, taken close to the Sun's opposition - since Gaia is in L$_2$ - we observe many asteroids: between $30$ and $100$ asteroids every night, up to magnitude $22$. In order to extract the astrometric positions as well as the magnitudes of these asteroids,  we have developed semi-automatic methods, strategies and tools tailored explicitly for this daily task. In only three and a half years of operation, this system has allowed us to send to the Minor Planet Center the position and the photometry for about $20,000$ asteroids, amongst which $9,000$ new objects. Here  we describe all the aspects of the GBOT asteroid survey.}

\resumen{El grupo GBOT (Ground Based Optical Tracking) es el encargado del seguimiento óptico desde tierra del satélite Gaia. Concretamente, desde el lanzamiento de Gaia, nuestra tarea consiste en tomar cada noche una secuencia de $10$ o $20$ imágenes de la nave cuando ésta se encuentra cerca del meridiano utilizando telescopios terrestres medianos. 
Principalmente, utilizamos el VLT Survey Telescope y el Liverpool Telescope.
Debido a que dichas imágenes son tomadas en la dirección opuesta del sol (Gaia se encuentra en L2), en ellas también podemos observar una considerable cantidad de asteroides; a saber, entre $30$ y $100$ asteriodes cada noche con una magnitud de hasta $22$. Para lograr obtener toda la información relevante relacionada con estos objetos y enseguida enviarla al Centro de Planetas Menores (Minor Planet Center) de manera diaria, el equipo GBOT ha desarrollado distintos métodos, estrategias y herramientas especialmente pensados para esta misión. Durante tres años y medio de operaciones, este sistema nos ha permitido enviar al MPC la posición y la fotometría de alrededor de $20~000$ asteroides, entre los cuales $9000$ nuevos objectos.
En seguida exponemos los distintos aspectos de este programa de observación de asteroides realizado por el grupo GBOT.}

\addkeyword{minor planets, asteroids}
\addkeyword{surveys}
\addkeyword{astrometry}
\addkeyword{image processing}

\begin{document}
\maketitle

\section{Introduction}
\label{sec:Intro}

This proceeding reports the talk given by S. Bouquillon in October 2018 for the 8\textsuperscript{th} ADeLA meeting (Tarija, Bolivia). It presents an overview of the status and results of the GBOT Asteroids Survey as it was at that time. 

The main characteristics of this asteroid survey is that the observations are taken near-opposition allowing accurate measurements of the absolute magnitudes of these objects and our capacity to detect faint moving objects (until the $r$-magnitude $22$) enabling us to discover a large number of asteroids.

In Section \ref{sec:GBOT}, we describe succinctly the GBOT project: the project's scientific rationale and its purpose are explained in Section \ref{subsec:GBOTproject} while the adopted observational strategy and the tools developed to reach the objectives are underlined in Section \ref{subsec:GBOTtools}. 

Section \ref{sec:AstSurvey} is dedicated to the GBOT Asteroid Survey. We first discuss the specific interests of the GBOT data for the asteroid studies (\ref{subsec:AstSurvey_Interest}), and describe our reduction procedures and analysis approach to extract the astrometry and photometry of these moving objects  (\ref{subsec:AstSurvey_Proc}). Finally,  in Section \ref{subsec:AstSurvey_Stat}, we present statistics on the collected data, as well as our open database system where all these information are save.

In Section \ref{sec:prospect}, we conclude by presenting our motivations for a second epoch follow-up within $24$ to $48$ hours from the first detection; as well as multi-band photometry which provide us with valuable information about the colours of these objects.

\section{The optical tracking of the Gaia satellite (GBOT)}
\label{sec:GBOT}

\subsection{The purposes of the GBOT project}
\label{subsec:GBOTproject}

The Gaia Ground Based Optical Tracking (\citet{2014SPIE.9149E..0PA}, hereafter GBOT) is a working group in the Gaia Data Processing and Analysis Consortium (DPAC) which gathers the scientists and the developers responsible for the processing of Gaia's data with the final objective of producing the Gaia Catalogues. 
The GBOT group is charged with obtaining optical positional measurements of the Gaia satellite to supplement the radiometric tracking in the determination of the satellite's orbit at the precision required for Gaia's scientific objectives.

For this purpose, GBOT has to make optical CCD observations of the Gaia satellite with ground based telescopes, perform astrometric reduction of these images, measure the position of Gaia in the ICRF with a $20$ mas daily accuracy and provide them to the European Space Operations Centre\footnote{An overview of the Gaia-GBOT project can be found on the GBOT web site \url{http://gbot.obspm.fr/}.}.

\subsection{The GBOT operations and tools}
\label{subsec:GBOTtools}

The observational strategy adopted consists in taking every night short sequences of images of the Gaia satellite close to its meridian transit time to minimize the airmass. We mainly use ESO’s  2.6\,m VLT Survey Telescope (VST Paranal, Chile), and the 2.0\,m Liverpool Telescope (LT La Palma, Spain). Occasionally, the two 2.0\,m Faulkes Telescopes in Australia and in Hawaii are also used.
The number of frames, the exposure times, the filters or the telescope tracking mode have been selected according to the characteristics of the target (the Gaia satellite) and the specificities of each telescope. 
Particular attention has been given to optimizing the astrometry of faint and fast objects such as Gaia which seen from Earth has a $r$-magnitude of $21.5$  and an angular motion that can be up to $2$”/min \citep{2017A&A...606A..27B}.

We adopted the following sequences of observations (all in SDSS-$r$ filter): $20$ frames of $60$ seconds with a sidereal tracking mode\footnote{This choice is dictated by technical limitation of the non-sidereal tracking mode of LT which is not completely even and smooth.} for the LT, $10$ frames of $60$ seconds with a tracking mode locked on Gaia for the VST.

The day-to-day GBOT operations proceed as follows: we prepare the observations of the following night, download images of the previous night, perform the reduction, measure Gaia's positions and store all the relevant information in a dedicated database. For all these steps, the GBOT group has developed its own procedure, analysis tools, and database system \citep{2014SPIE.9152E..03B}. In particular, the \emph{GBOT Astrometric Reduction Pipeline} - a fortran-based program for CCD-image reduction - has been elaborated and optimized for the astrometry of moving objects, (publically available from \url{http://gbot.obspm.fr/pipeline}).

\section{The GBOT Asteroid Survey}
\label{sec:AstSurvey}

\subsection{Benefit of GBOT data for Asteroids observations and science}
\label{subsec:AstSurvey_Interest}

\begin{figure}[!h]\centering
  \includegraphics[width=0.7\columnwidth]{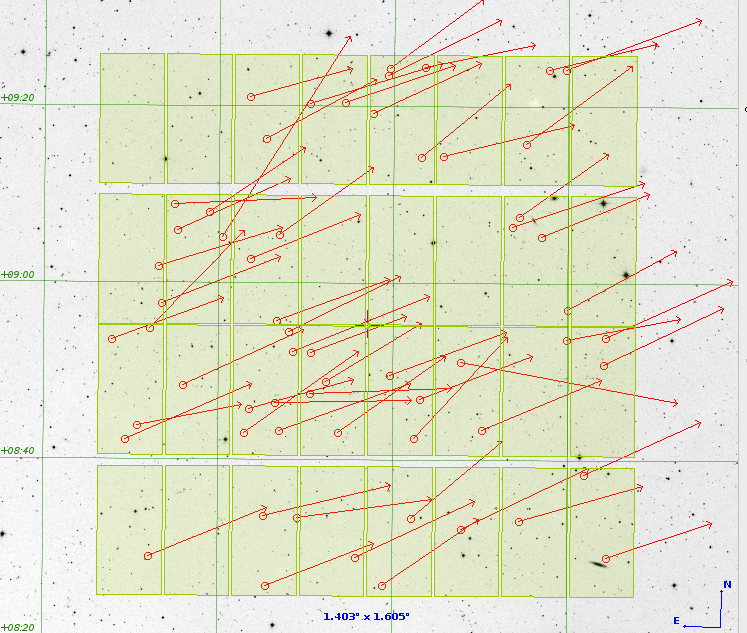}
  \caption{Asteroids observed on March 18\textsuperscript{th}, 2015. The red circles and arrows show the positions and velocities of the detected asteroids, while the green shaded area is the field of view covered by OmegaCAM superposed to a DSS image with Aladin software \citep{2011ASPC..442..683B}.}
  \label{fig:fov}
\end{figure}

Each day, some asteroids are visible close to the line of sight of Gaia and are serendipitously recorded in the frames of the GBOT sequences. For instance, as it can be seen in Figure \ref{fig:fov}, $60$ asteroids passed through the field of view of the GBOT images taken with VST on March 18\textsuperscript{th}, 2015.

Out of the $60$ asteroids detected during this night, $33$ were linked to known asteroids existing in the Minor Planet Center (MPC) Database while the remaining ($27$) are unknown or known objects with inaccurate orbital parameters at the observation date.

This significant number of asteroids present in the GBOT images are due to the the fact that, as we observe Gaia which is in the Lagrangian point Sun-Earth L$_2$, the GBOT images are close to the Sun's opposition. This means that we observe in the vicinity of the celestial ecliptic plane where the majority of asteroids are and in an area where they are brighter (at opposition, the Earth-asteroid distance is the shortest and their phase angle is zero). The large field of view 
of OmegaCAM (the VST camera of one square degree) as well as the necessity for the Gaia tracking to be able to record faint moving objects until $r$-magnitude $22$ are other reasons which explain the large amount of detections.  

The scientific interests of the asteroids observed by GBOT is twofold. 
First, as these objects are observed near-opposition, their absolute magnitudes\footnote{For asteroids, the absolute magnitude is defined as the brightness, normalzed at 1 AU distance from both Sun and Earth, that the object would have if it was seen exactly at opposition} can be measured accurately. This parameter, that Gaia or \emph{WISE} asteroid Surveys are unable to measure\footnote{Gaia and \emph{WISE} satellites which operate only at limited inclinations with respect to the sun, can not observe asteroids near-opposition.}, is important since is directly related to the asteroids surface's properties via their albedos.
Secondly, our capacity to observe and detect faint moving objects until the $r$-magnitude $22$, which is quite unusual for standard asteroid surveys, gives us the opportunity to discover a large number of asteroids.

\subsection{GBOT Asteroids: processing the data}
\label{subsec:AstSurvey_Proc}

The GBOT system for detecting and extracting the astrometry and photometry of asteroids is operational since January 2015 (one year after the launch of Gaia). The daily GBOT asteroid operations are organised as follows: 
\begin{compactitem}
\item The first step, is the astrometric and photometric calibration of all CCDs of OmegaCAM and not only the CCD where Gaia is. This is performed with the GBOT reduction pipeline in the same way as before.

\item The second step is the detection of objects moving in the GBOT sequences. This, is performed by an upgraded version of our Fortran-code named \emph{TargetFinder} originally in charge of the Gaia detection and based on a simple \emph{Cone Search} \citep{2014SPIE.9152E..03B}. The new method implemented is based on Hough Transformation (HT), a well-known method for moving objects detection in Computer Vision \citep{Illingworth1988} but rarely used for astronomical application. 

\item The third step establishes the link between the detected asteroids and the known ones in the MPC database.
This is done by the integration of \emph{Skybot}\footnote{\citep{2006ASPC..351..367B}} request capabilities into the \emph{TargetFinder} code. 
 \end{compactitem}
 
The data relative to all detected asteroids are then gathered and are validated manually. This validation is required as the two automatic processes - asteroid detection and identification - may result in a  small but non-negligible number of erroneous detections, e.g: false detections in the diffraction and reflection patterns of bright stars can be identified as faint moving objects by our HT code, a faint star not in the reference catalogue can overlap with one asteroid in one image, etc.

 Once this last step has been performed, the relevant data concerning all the detected asteroids are
stored in the GBOT Asteroid Archive\footnote{\url{http://gbot.obspm.fr/asteroids/}} and their astrometry sent to MPC less than $15$ hours after their observations helping the follow-up of the newly discovered objects by providing a first astrometric position and allowing the improvement of the orbit parameters of the known objects. 

\subsection{GBOT asteroids: statistics and archive}
\label{subsec:AstSurvey_Stat}
Here we discuss the GBOT asteroids data gathered between January 1\textsuperscript{st}, 2015 and May 15\textsuperscript{th}, 2018, with the two main GBOT sequences of images taken with VST and LT.

\begin{figure}[!h]
  \includegraphics[width=\columnwidth]{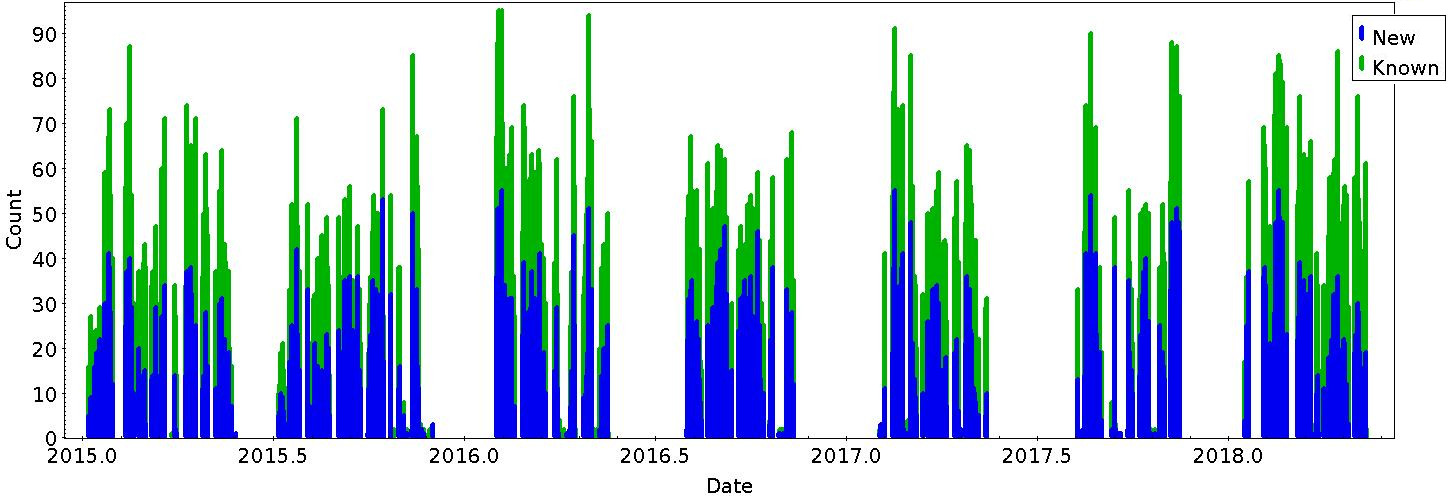}
  \caption{Number of asteroids detected per night between January 2015 and June 2018. The known objects are represented in green, whereas the newly discovered are represented in blue.
  \label{fig:DetperNight}}
\end{figure}

On average between $30$ to $100$ candidate asteroids were detected for each night (as can be seen in Figure~\ref{fig:DetperNight}).

A total of $18,948$ asteroids were detected, of which $9,943$ were matched to MPC entries and $9, 005$ were not matched (at the observation date). A large majority of these asteroids have been observed with VST due to its wide field of view and its larger mirror size. However a non-negligible number of them have been observed with LT ($289$ known objects and $218$ new objects respectively).  

In Table \ref{tab:AstNbperGroup} the classification by dynamical groups of the previously known objects re-observed by GBOT are shown. We see that most of the detected objects are MBAs (Main-Belt Asteroids).

\begin{table}\centering
  \setlength{\tabnotewidth}{\columnwidth}
  \tablecols{2}
  \setlength{\tabcolsep}{2\tabcolsep}
  \caption{Number of asteroids observed by GBOT according to their groups} 
  \label{tab:AstNbperGroup}
  \begin{tabular}{lr}
    \toprule
Group & \multicolumn{1}{c}{Number of asteroids} \\
    \midrule
    Near Earth (Amor)   &  	$9$   \\	
    Near Earth (Apollo) & 	   $10$    \\ 	
    Hungaria     &	    $32$   \\
    Mars-Crosser &  	$58$   \\
    Main Belt (Inner)   &	    $3660$ \\
    Main Belt (Middle)   &	    $3351$ \\
    Main Belt (Outer)   &	    $2657$ \\
    Main Belt (Cybele) & $30$ \\
    Main Belt (Hilda) & $61$ \\
    Trojan   	 &	    $73$ 	 \\
    Centaur   	 &  	$2$ 	 \\ 
    \bottomrule    
  \end{tabular}
\end{table}

\begin{figure}[!h]
  \includegraphics[width=\columnwidth]{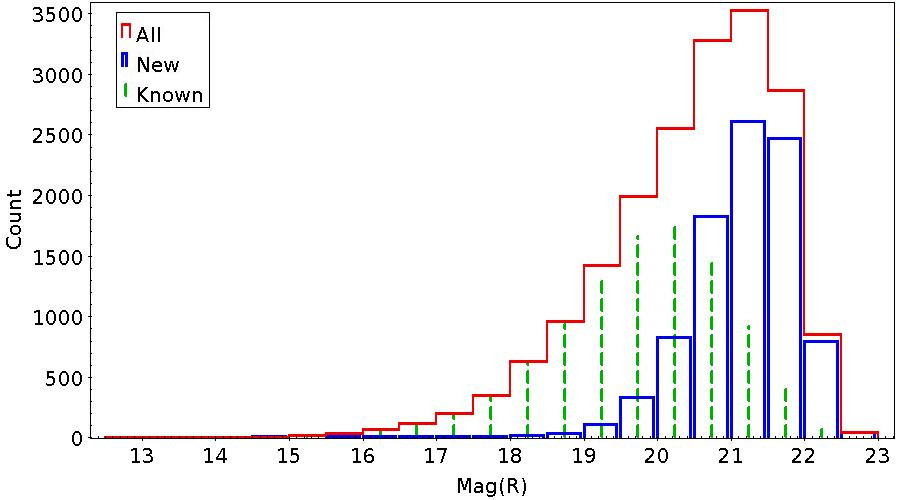}
  \caption{Cumulative number of asteroids detected as a function of the measured $r$-magnitude over the period Jan. 1\textsuperscript{st}, 2015 - May 15\textsuperscript{th}, 2018. We represent in green the known objects, in blue the objects that were unknown at the time of the observation, and  in red the grand total. 
 \label{fig:mag}}
\end{figure}

Figure \ref{fig:mag} shows the number of cumulative observed asteroids as a function of the measured \mbox{$r$-magnitude} (SDSS). We detect a wide range of magnitude, going from $13$~mag. for the brightest up to mag.~$23$ for the faintest. Most of the detected objects have magnitudes in the range $19$ to $22$.

\begin{figure}
\begin{center}
  \includegraphics[width=\columnwidth]{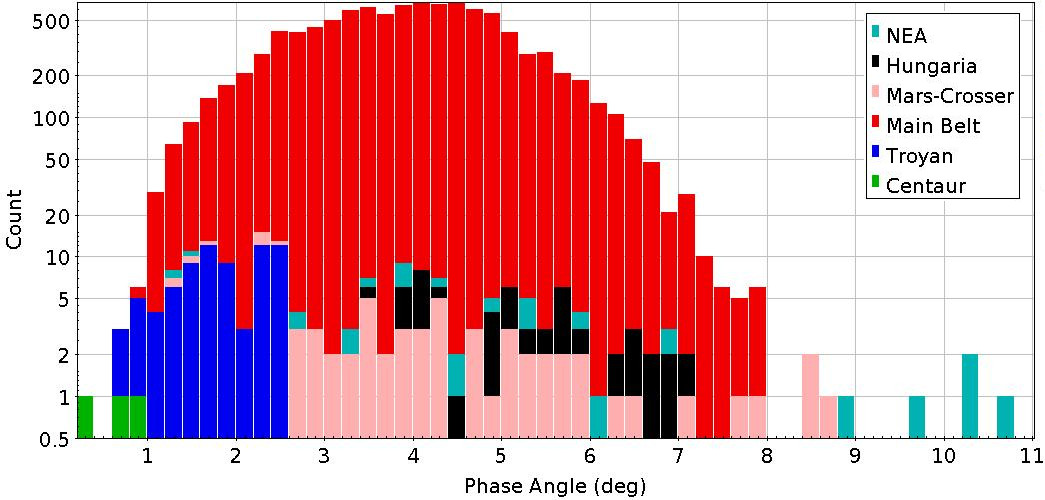}
  \caption{Cumulative histogram of the known asteroids (per dynamical population) detected by GBOT as a function of the phase angle over the period Jan 1\textsuperscript{st}, 2015 - May 15\textsuperscript{th}, 2018.
  \label{fig:PhAngle}}
\end{center}
\end{figure}

As explained in previous section, all the detected objects are observed near-opposition. Figure \ref{fig:PhAngle} shows the number of discovered asteroids as a function of the phase angle. On average the asteroids are observed with a phase angle $\sim4^\circ$.

\section{Conclusion \& Prospects}
\label{sec:prospect}
The GBOT Asteroids Survey is a unique program that exploits the optical tracking observations of Gaia to detect asteroids close to opposition.

The Survey as presented here covers the Jan. 15\textsuperscript{th}, 2015 to May 15\textsuperscript{th}, 2018 period. It has allowed to improve the astrometry of $9,943$ known asteroids and the detection of $9,005$ asteroids unknown at the observation date, only $2,322$ ($\sim25\%$) are now recorded in the MPC database, as of April 2020. This means that $\sim75\%$ of them ($6,683$ objects) have never been re-observed or identified and are currently lost. Furthermore, these objects were observed in only one colour (SDSS-$r$) which does not allow us to do any taxonomic classification.

In 2018 we added to the GBOT Asteroids Survey two follow-up 24 or 48 hours after the first detection: a first systematic one  since April with VST of all the detected asteroids in 4 different filters SDSS (g,r,i,z) and a second one since December with Las Cumbres Observatory telescope network for selected asteroids with SDSS-$r$ filter. These two follow-up will reinforce the scientific value by: 

\begin{compactitem}
\item providing another astrometric observation, reducing the loss of new unconfirmed objects,

\item obtaining a preliminary taxonomic classification to estimate the surface composition of these objects.

\item deriving from the multi-color photometry the Johnson V absolute magnitudes ($H$) and the slope parameters ($G$) , which describes the shape of the magnitude phase function.   
\end{compactitem}

\end{document}